\documentstyle[aps,prl,multicol,epsf]{revtex} 
\begin{document}
\def\B.#1{{\bbox{#1}}}
\def\BC.#1{{\bbox{\cal{#1}}}}
\title{{\rm PHYSICAL REVIEW LETTERS \hfill \sl Submitted}\\
Fusion Rules in Navier-Stokes Turbulence:
 First Experimental Tests} 
\author {*Adrienne L. Fairhall, \dag Brindesh Dhruva,
 *Victor S. L'vov, 
*Itamar Procaccia and \dag Katepalli R. Sreenivasan}
\address{*Department
of~~Chemical Physics, The Weizmann Institute of Science,
 Rehovot 76100, Israel\\ 
 \dag Mason Laboratory, Yale University, New Haven, CT 06520-8186}
\maketitle
\begin{abstract}
We present the first experimental tests of the recently derived fusion
rules for Navier-Stokes (N-S) turbulence. The fusion rules address the
asymptotic properties of many-point correlation functions as some of
the coordinates coalesce, and form an important ingredient of the
nonperturbative statistical theory of turbulence.  Here we test the
fusion rules when the spatial separations lie within the inertial
range, and find good agreement between experiment and theory. An
unexpected result is a simple linear law for the Laplacian of the
velocity fluctuation conditioned on velocity increments across large
separations.
\end{abstract}
\begin{multicols}{2}

In this Letter the predictions of the recently proposed fusion rules
\cite{LP-1} are tested by analyzing the turbulent velocity signal at
high Reynolds number. We start with a short theoretical summary.

The theory focuses on two-point velocity differences
\begin{equation}
{\B.w}({\B.r},{\B.r}',t) \equiv {\B.u} (\B.r',t)- {\B.u}(\B.r,t), 
\label{du1}
\end{equation}
where ${\B.u} (\B.r',t)$ is the Eulerian velocity field accessible to
experiment.  One attempts to extract predictable and computable
results by considering the statistical properties of $\B.w$
\cite{MY,Fri}. The most informative statistical quantities are the
equal-time rank-$n$ tensor correlation functions of velocity
differences \begin{eqnarray} &&\B.{\cal
F}_n({\B.r}_1,{\B.r}'_1;{\B.r}_2,{\B.r}'_2; \dots;{\B.r}_n,{\B.r}'_n)
\nonumber \\
&=& \langle \B.w({\B.r}_1,{\B.r}'_1)
\B.w({\B.r}_2,{\B.r}'_2) \dots
\B.w({\B.r}_n,{\B.r}'_n) \rangle \ ,
\label{defF}
\end{eqnarray}
where $\langle \cdot \rangle$ denotes averaging, and all coordinates
are distinct.  We consider N-S turbulence for which the scaling
exponents are presumed to be universal (i.e., they do not depend on
the detailed form of forcing), and where the correlations $\B.{\cal
F}_{n}$ are homogeneous functions \cite{MY}, namely
\begin{equation}
\B.{\cal F}_{n}(\lambda{\bbox r}_1,
\lambda{\bbox r}'_1,\!\dots\! ,\lambda{\bbox r}'_{n})
=\lambda^{\zeta_{n}}\!\B.{\cal F}_{n}
({\bbox r}_1,{\bbox r}'_1, \!\dots\!,{\bbox r}'_{n}), 
\label{hom}
\end{equation}
$\zeta_n$ being the homogeneity (or scaling) exponent. This form
applies whenever all the distances $\left|\B.r_i-\B.r'_i\right|$ are
in the ``inertial range", i.e.  between the outer scale $L$ and the
dissipative scale $\eta$ of the system. Our aim here is to describe
the behavior of such functions as pairs of coordinates approach one
another, or ``fuse''. The fusion rules, derived in \cite{LP-1,LP-2},
govern the analytical structure of the correlation functions under
this coalescence.

The statistical function that has been most commonly studied
\cite{MY,Fri,84AGHA,Sre} is the structure function $S_n(R)$
\begin{equation} S_n(R) = \left\langle\vert\ {\B.w}({\B.r},{\B.r}')
\vert^n\right\rangle,
\quad \B.R\equiv {\B.r}'-{\B.r}. \end{equation}

Clearly the structure function is obtained from (\ref{defF}) by the
fusing of all coordinates $\B.r_i$ into one point $\B.r$, and all
coordinates $\B.r'_i$ into $\B.r'=\B.r+\B.R$. In doing so, one crosses
the viscous dissipation length-scale. One then expects a change of
behavior, reflecting the role of the viscosity in the theory for
$S_n(R)$. In developing a N-S based theory in terms of $S_n(R)$, one
encounters the notorious closure problem: one must balance terms
arising from the convective term $\B.u \cdot \B.\nabla\B.u$ and the
dissipative $\nu\nabla^2 \B.u$ term, neither of which can be
neglected. Hence determining $S_n(R)$ requires information about
$S_{n+1}(R)$. All known closures of this hierarchy of equations are
arbitrary. However, according to the theory of Refs. \cite{LP-1,LP-2},
the fully unfused ${\BC.F}_n$ does not suffer from this problem. When
all separations are in the inertial range, the viscous term may be
neglected, and one obtains \cite{LP-2} homogeneous equations for
${\BC.F}_n$ in terms of ${\BC.F}_n$ only, with no hierarchic
connections to higher or lower order correlations. Such homogeneous
equations may exhibit new, anomalous scaling solutions for the
correlation functions ${\BC.F}_n$.

There are various possible configurations of coalescence. We will test
only those fusions in which the coalescing points are those of
velocity differences. One can also consider the coalescence of points
from different velocity differences, but they are experimentally more
difficult to measure; we will ``precoalesce'' all such points here and
comment on its effects \cite{LP-2}. The first set of fusion rules that
we examine concerns ${\BC.F}_n$ when $p$ pairs of coordinates
$\B.r_1,\B.r'_1\dots \B.r_p,\B.r'_p$, $(p<n)$ of $p$ velocity
differences coalesce, with typical separations between the coordinates
$|\B.r_i -\B.r'_i| \sim r$ for $i \le p$, and all other separations of
the order of $R$, $r\ll R \ll L$.  The fusion rules predict
\begin{eqnarray}\label{fusion1}
&& {\BC.F}_n({\B.r}_1,{\B.r}'_1;
\dots;{\B.r}_n,{\B.r}'_n)\\
&=&
{\tilde{\BC.F}}_p({\B.r}_1,{\B.r}'_1;
\dots;{\B.r}_p,{\B.r}'_p)
\bbox\Psi_{n,p}
({\B.r}_{p+1},{\B.r}'_{p+1};\dots;{\B.r}_n,{\B.r}'_n) \ , 
\nonumber
\end{eqnarray}
where ${\tilde{\BC.F}}_p$ is a tensor of rank $p$ associated with the
first $p$ tensor indices of ${\BC.F}_n$, and it has a homogeneity
exponent $\zeta_p$.  The $(n-p)$-rank tensor
$\bbox\Psi_{n,p}({\B.r}_{p+1},{\B.r}'_{p+1};\dots;{\B.r}_n,{\bf
r}'_n)$ is a homogeneous~~function~~with~~a~~scaling\hfill  exponent

\narrowtext
\begin{figure}
\epsfxsize=8.5truecm
\caption
{Log-log plot of the structure functions $S_n(R)$ as a function of $R$
for $n=2,4,6,8$ denoted by +, $\times$, $*$ and $\circ$ respectively. }
\label{Fig1}
\end{figure}
\noindent
$ \zeta_n-\zeta_p$, and is associated with the remaining $n-p$ indices
of ${\BC.F}_n$.  The case $p=1$ is special, since the leading order
evaluation cancels by symmetry.  The next-order result, for a randomly
oriented set of pairs with the separation $R$, is
\begin{equation}
{\BC.F}_n({\B.r}_1,{\B.r}'_1;\dots;{\B.r}_n,{\B.r}'_n) \sim (r/ R)
S_n(R). \
\label{rule2}
\end{equation}

One can also consider correlation functions under the operation of
gradients.  One such set of functions, denoted $J_n(R)$, have
particular significance.  They arise from the dissipative term when
one obtains a statistical balance from the N-S equations. For
comparison with experiments we define $J_n(R)$ in terms of the
longitudinal velocity difference $\delta u_R \equiv
{\B.w}(\B.r,\B.r+\B.R)\cdot\B.R/R$ as
\begin{equation}
J_n(R)=\nu \left<\tilde \nabla^2\B.u(\B.r) 
[\delta u_R]^{n-1}\right>. \label{Jn}
\end{equation}
The Laplacian operator in (\ref{Jn}) is interpreted as a finite
difference of longitudinal components of the velocity,
\begin{equation}
\tilde \nabla^2\B.u(\B.r) = [{\B.w}(\B.r,\B.r
+\B.\rho)- {\B.w}(\B.r,\B.r-\B.\rho)]
\cdot\B.\rho/\rho^3. \label{surr} \end{equation}
The predictions of the fusion rules are different for $\rho$ above and
below the dissipative scale. They read:
\begin{eqnarray}
J_n(R)&=&nC_nJ_2 S_n(R)/2S_2(R) \ , \quad \rho\gg\eta\ , 
\label{bigrho}\\
J_n(R)&=&n\tilde C_n J_2 S_{n+1}(R)/S_3(R) \ , \quad 
\rho\ll\eta \ , \label{smallrho}
\end{eqnarray}
where $C_n$ and $\tilde C_n$ are $R$-independent dimensionless
constants.  The fusion rules do not rule out an $n$-dependence of
these coefficients. $J_2$ is expected to be $R$-independent.

Here we will test these predictions using atmospheric turbulence data
obtained by means of a single hot-wire probe mounted at a height of 35
m on the meteorological tower at the Brookhaven National
Laboratory. The hot-wire was about 0.7 mm in length and 6 $\mu$m in
diameter. It was operated on a DISA 55M01 anemometer in the constant
temperature mode. The wind direction, measured independently by a vane
anemometer, was approximately constant. The frequency response of the
hot-wire was 
good up to 5kHz. The voltage from the anemometer were
low-pass filtered at 2 kHz and sampled at 5 kHz, and later converted
to velocity through an in-situ calibration. The mean wind speed was
7.6 ms$^{-1}$ and the root-mean-sqaure velocity was 1.3
ms$^{-1}$. With the usual procedure of surrogating time for space
(``Taylor's hypothesis"), we obtain the Taylor microscale Reynolds
number to be 9540 and the Kolmogorov microscale to be 0.57 mm.

In Fig.~\ref{Fig1} we present the structure functions $S_n(R)$ as a
function of $R$.  They were computed using 10 million data samples. In
this and other figures, spatial separations have units of sampling
times, and the velocity is normalised by the RMS velocity.  
This figure shows that we have around three decades of
``inertial range'' (between, say 10 and $10^4$ sampling units) but
that the dissipative range is not well-resolved. Structure functions
of orders higher than 8 are less reliable and will not be considered.

While the fusion rules are formulated for differences in
$d$-dimensional space, the surrogated data represent a 1-dimensional
cut. This has implications for the choice of positioning of the
coordinates. In $d$-dimensional space we can choose separations to
fall within balls of size $R$ and $r$ respectively.  In our case this
ball collapses onto a line, and best results are obtained when the
pairs of coordinates in the two groups coincide. As a simple
demonstration consider the second order quantity ${\cal
F}_2(\B.r_1,\B.r'_1,\B.r_2,\B.r'_2)$ with the three different choices:
(i) $r_1=r'_1=x$, $r_2=r'_2=y$, $r=|y-x|$, (ii) $|r_1-r'_1|=
|r_2-r'_2|=r$, $r'_1=r_2$, (iii) $|r_1-r'_1|=|r_2-r'_2|=r$ where
$r'_1$ and $r_2$ are also separated by $r$. In one dimension one can
simply compute the ratio of the correlation functions in cases (ii)
and (iii) with respect to case (i). One finds a reduction factor
$2^{\zeta_2-1}-1\approx -0.2$ and
$(-2^{\zeta_2+1}+1+3^{\zeta_2})/2\approx -0.05$. We thus see that
there is a rapid decrease in amplitude when the distances are not
enmeshed, and so all averaging is done using maximally enmeshed
configurations (i.e., case (i)). As remarked above, in doing so, all
the ``unprimed" points {\em not across velocity differences} are
already fused. This procedure does not affect predictions
(\ref{fusion1}) and (\ref{rule2}), see \cite{LP-2}.

Explicitly, therefore, we examine the behavior of the correlation
function
\FL
\begin{equation}
{\cal F}_{p+q}(r,R)\!
\equiv \!\langle [u(x+r)\! - \!u(x)]^p [u(x+R)\! -\! u(x)]^q 
\rangle \label{partfuse}
\end{equation}
as a function of both $r$ and $R$ for several values of the powers $p$
and $q$. From Eq.~(\ref{fusion1}) one expects \begin{equation} {\cal
F}_{p+q}(r,R) \sim S_p(r) S_{q+p}(R)/ S_p(R).
\label{fusewham}
\end{equation}

\narrowtext
\begin{figure}
\epsfxsize=8.0truecm
\caption
{Log-log plot of ${\cal F}_{p+q}(r,R) $ as a function of $r$ at fixed
$R$ for $q=2$ and $p=2,4,6$ denoted by $+$, $\times$ and $*$
respectively with dashed lines.  Shown with dotted lines are the same
quantities divided by $S_p(r)$.}
\label{Fig2}
\end{figure}
\begin{figure}
\epsfxsize=8.1truecm
\caption
{As in Fig.~2 as a function of $R$ at fixed $r$, with the dotted lines 
representing the quantities divided by $S_{p+q}(R)/S_p(R)$.}
\label{Fig3}
\end{figure}
\narrowtext
\begin{figure}
\epsfxsize=8.0truecm
\caption
{Log-log plot of ${\cal F}_{1+q}(r,R) $ as a function of $R$ for
$q=1,3,5$ denoted by +, $\times$ and $*$ respectively. Shown with
dotted lines are the same quantities divided by $S_{1+q}(R)/R$.}
\label{Fig4}
\end{figure}

\begin{figure}
\epsfxsize=8.0truecm
\caption
{$\log_{10}[J_n(R)]$ as a function of the fusion rule prediction
$\log_{10}[J_2S_n(R)/S_2(R)]$ for $n=2,4,6,8$ denoted by +, $\times$,
$*$, $\circ$ respectively. Inset: the coefficient $C_n$ with the same
notation. }

\label{Fig5}
\end{figure}

\noindent
In Fig.~\ref{Fig2} we display the results for $q = 2$ with even values
of $p$ as a function of $r$ for $r$ in the inertial range. Only even
values are displayed as the odd correlations fluctuate in sign.  The
large scale $R$ was fixed at the upper end of the inertial range. The
data show clean scaling in the inertial range. Overlaid are the
averages corrected by the prediction of the fusion rule
(\ref{fusewham}). Here and in all other figures the averages
themeslves are connected with dashed lines, whereas compensated
results are shown dotted. One observes a change of behavior as $r$
approaches $R$ at the upper limits and the average increases in size
towards the ``fully-fused'' quantity, $S_{p+q}(R)$. Similarly
convincing results were obtained for other values of $p$ and $q$.

In Fig.~\ref{Fig3} we show ${\cal F}_{p+q}(r,R)$ as a function of the
large scale $R$ with the small separation $r$ fixed at the low end of
the inertial range, together with the values corrected by
(\ref{fusewham}).  There is a clear trend towards zero slope in the
corrected quantity in the upper inertial range.

We consider now the special case that a single pair of points in a
velocity difference approach one another. The prediction of
Eq.~(\ref{rule2}) is tested for fixed $r$ and the expected dependence
on $R$ is well-verified in Fig.~\ref{Fig4}.  The results found by
varying $r$ are not shown here: the linear configuration of all the
measurement points leads to a competition between the leading and next
order of scaling.

The function $J_2$ was computed and confirmed to be constant
throughout the inertial range. In Fig.~\ref{Fig5} we present $J_n(R)$
as a function of $nJ_2 S_n(R)/2S_2(R)$ for $n$, $n=2,4,6,8$ and
inertial range $R$. The finite difference surrogate of the Laplacian
(cf. Eq. (\ref{surr})) was computed with $\rho=10$. The straight line
$y = x$ passing through the data is not a fit. It appears from these
results that (\ref{bigrho}) is obeyed well with $C_n=1$. The $R$
independence of $C_n$ is 

\narrowtext
\begin{figure}
\epsfysize=7truecm
\epsfxsize=8.0truecm
\caption{The conditional averages of the Laplacian} 
\label{Fig6}
\end{figure}
\noindent
a direct confirmation of the fusion rules for
the fusion of two points. On the other hand, the fact that $C_n$ is
independent of $n$ is a surprise that does not follow from fusion
rules, and has interesting implications for the statistical theory of
turbulence. A more sensitive check of the value of $C_n$ is obtained
by dividing $nJ_2S_n(R)/S_2(R)$ by $J_n(R)$ for individual values of
$n$ and $R$. This is displayed in the inset in Fig.~5. Clearly, there
are statistical fluctuations that increase with increasing $n$ but the
data show that $C_n$ is approximately constant in $R$ and $n$, with a
value of about unity.

If indeed $C_n$ is $n$-independent, there are surprising consequences 
for the conditional statistics of our field. To see this, $J_n(R)$ may
be rewritten
\begin{equation}
J_n(R) = \int d\delta u_R P[\delta u_R]
\langle\tilde \nabla^2\B.u(\B.r)\vert 
\delta u_R\rangle \delta u_R^{n-1} \ . \label{int}
\end{equation}
Here $\langle\tilde \nabla^2\B.u(\B.r)\vert \delta u_R\rangle$ is the
average of the finite difference Laplacian conditioned on a value of
$\delta u_R$. The only way of satisfying both (\ref{int}) and
(\ref{bigrho}) with $C_n$ that is independent of $n$ and $R$ is to
assert that the conditional average, which is in general a function of
$ \delta u_R $ and $R$, can be factored into a function of $R$ and a
linear function of $\delta u_R$:
\begin{equation}
\langle\nabla^2\B.u(\B.r)\vert \delta u_R\rangle = 
{J_2 \over 2S_2(R)} \delta u_R.
\label{linear}
\end{equation}

Such linear laws have been discussed in the context of conditional
statistics of passive scalar advection 
\cite{SY,KSS,94Kra,96FGLP,96CLPP}, and were thought to be reasonable 
because of the linear nature of the advection-diffusion equation for
the scalar. Thus, linear laws for N-S turbulence as well were
unexpected. In Fig.~6 we display a direct calculation of the
conditional average of the surrogate Laplacian with $\rho=10$,
multiplied by $2S_2(R)/J_2$ as a function of $\delta u$ for nine
values of $R$ spanning the inertial range.  All the data collapse on a
single straight line whose slope is unity. For ease of presentation we
have displaced the different values of $R$ from each other. It appears
that the prediction (\ref{linear}) is amply supported by the data.

Unfortunately we cannot test (\ref{smallrho}) with the present data
because sub-dissipation scales are not resolved. The prediction
implies that the nature of the conditional average changes
qualitatively when $\rho$ decreases below the dissipative scale. Such
changes have important consequences for the ultraviolet properties of
the statistical theory of turbulence, and a rich variety of
predictions are already available \cite{LP-3}. It is thus worthwhile
generating high-Reynolds-number data that resolve sub-dissipative
scales. Efforts to acquire such data are under way.

This work was supported in part by the German Israeli Foundation, the
US-Israel Bi-National Science Foundation, the Minerva Center for
Nonlinear Physics, and the Naftali and Anna Backenroth-Bronicki Fund
for Research in Chaos and Complexity.  We thank the Brookhaven
National Laboratory for permission to use their facilities and
Mr. Victor Cassella for his help in setting up the experiment.

\end{multicols}
\end{document}